# SYSTEMATIC EFFECTS IN PROPER MOTION OF RADIO SOURCES

O. Titov
Geoscience Australia
PO Box 378 Canberra 2601 Australia
e-mail: oleg.titov@ga.gov.au

**Abstract**

.The galactocentric rotation of the Solar system generates the systematic effect in proper motions known as 'secular aberration drift'. This tiny effect (about five microseconds per year) in the quasar proper motion can be measured by VLBI. However, the motions of relativistic jets from the active extragalactic nuclei can reach several hundred microseconds per year and mimic the proper motion of the observed radio sources. These apparent motions exceed the secular aberration drift by factor of 10-100. In this paper we search the ways to overcome the difficulties and discuss our estimates of the secular aberration drift using OCCAM software.

## 1. Introduction

Multifrequency VLBI techniques measure accurate positions of the reference extragalactic radio sources. From analysis of a global set of VLBI data the positions are estimated with accuracy up to 0.1 mas for frequently observed radio sources (more than 10,000 observations per source). However, the effect of the intrinsic sources structure causes linear and non-linear variations in the sources astrometric positions (Ma et al, 1998).The apparent propermotion can reach several hundred microarcsec/year for some radio sources. The positional variations are supposed to be uncorrelated over the sky, and do not produce any systematic pattern. Nonetheless, some systematic effects in the apparent radio sources proper motions has been predicted. The galactocentric acceleration of the Solar system barycentre should cause the first order electric type vector spherical harmonic of the magnitude 4-5 microarcsec/year (Gwinn et al., 1997; Sovers et al., 1998; Kovalevsky, 2003; Klioner, 2003; Kopeikin, 2005). Primordial gravitational waves in the early Universe would produce the second order vector spherical harmonics of electric and magnetic types (Gwinn et al., 1997; Pyne et al., 1996). More generalized expression for the proper motion in a frame of the general relativity has been published by Kristian and Sachs (1966). The authors found that in the expanding Universe the proper motion of distant objects could increase with the distance.

Some attempts to detect the systematic effects have been done by different authors (Gwinn et. al., 1997; MacMillan, 2003; Titov, 2008). Gwinn el al. (1997) did not find any systematic in the proper motions using data from 1979 till 1996. MacMillan (2003) used VLBI data from 1979 till 2002 and declared detection of some systematics though without any details. Titov (2008) found some statistically significant amplitude of the first and second degree electric type harmonics using data from 1980 till 2007.

This paper presents more rigorous results than (Titov, 2008) on the systematic effects in the radio source proper motion. This includes a separate consideration of the vector spherical harmonics of the first and second order. The second degree 'magnetic' spherical harmonics were added to the list of estimated parameters. More red shifts of the reference radio sources are available from the database recently developed by Malkin and Titov (2008).

## 2. Vector spherical harmonics

Let us consider $\bar{F}(\alpha,\delta)$ as a vector field of a sphere described by the components of the proper motion vector ($\mu_\alpha \cos\delta, \mu_\delta$)

$$\bar{F}(\alpha,\delta) = \mu_\alpha \cos\delta \cdot \bar{e}_\alpha + \mu_\delta \cdot \bar{e}_\delta$$

where $\bar{e}_\alpha, \bar{e}_\delta$ - unit vectors. A vector field of spherical harmonics $\bar{F}(\alpha,\delta)$ can be approximated by the vector spherical functions as follows

$$\bar{F}(\alpha,\delta) = \sum_{l=1}^{\infty} \sum_{m=-l}^{l} (a_{l,m}^E \bar{Y}_{l,m}^E + a_{l,m}^M \bar{Y}_{l,m}^M)$$

where $\bar{Y}_{l,m}^E, \bar{Y}_{l,m}^M$ - the 'electric' and 'magnetic' transverse vector spherical functions, respectively

$$\bar{Y}_{l,m}^E = \frac{1}{\sqrt{l(l+1)}} \left( \frac{\partial V_{lm}(\alpha,\delta)}{\partial \delta} \bar{e}_\alpha + \frac{\partial V_{lm}(\alpha,\delta)}{\partial \alpha \cos\delta} \bar{e}_\delta \right)$$

$$\bar{Y}_{l,m}^M = \frac{1}{\sqrt{l(l+1)}} \left( \frac{\partial V_{lm}(\alpha,\delta)}{\partial \alpha \cos\delta} \bar{e}_\alpha - \frac{\partial V_{lm}(\alpha,\delta)}{\partial \delta} \bar{e}_\delta \right)$$

The function $V_{l,m}(\alpha,\delta)$ is given by

$$V_{l,m}(\alpha,\delta) = (-1)^m \sqrt{\frac{(2l+1)(l-m)!}{4\pi(l+m)!}} P_l^m(\sin\delta) \exp(im\alpha)$$

where $P_l^m(\sin\delta)$ - the associated Legendre functions

The coefficients of expansion (8) $a_{l,m}^E, a_{l,m}^M$ to be estimated as follows

$$a_{l,m}^E = \int_0^{2\pi} \int_{-\pi/2}^{\pi/2} \bar{F}(\alpha,\delta) \bar{Y}_{l,m}^E {}^*(\alpha,\delta) \cos\delta d\alpha d\delta$$

$$a_{l,m}^M = \int_0^{2\pi} \int_{-\pi/2}^{\pi/2} \bar{F}(\alpha,\delta) \bar{Y}_{l,m}^M {}^*(\alpha,\delta) \cos\delta d\alpha d\delta$$

where * means a complex conjugation. This system of equations can be solved by the least squares method. In this research the coefficients are estimated as global parameters from a large set of VLBI data.

OBSERVATIONS AND DISCUSSION OF RESULTS

The first and second degree spherical harmonics were estimated by the least squares collocation method Titov (2004). The database comprises of about 3.9 million observations of group delay with different baseline and sources made in 3554 24-hour sessions between April, 1980 and April, 2007. The equatorial coordinates of more than 2000 radio sources were observed as global or 'arc' parameters (see solution description below). The Earth orientation parameters, correction to IAU2000 nutation offsets as well as station coordinates were estimated as 'arc' parameters. NNR and NNT constraints imposed the station positions for each 24-hour session. Clock offsets, troposphere wet delays and north-south and east-west gradients were estimated as stochastic parameters for each observational epoch. The vector spherical harmonics were treated as global parameters, similar to the approached used by MacMillan (2003}.

The first solution was based on the all radio sources observed in geodetic and astrometric VLBI programs without separation into 'stable' and 'unstable'. In the second and third solutions were used not 'unstable' radio sources for the reference frame according to the Feissel-Vernier classification (Feissel-Vernier, 2003). The number of the radio sources with measured red shift has increased significantly since the previous publication (Titov, 2008}. The second solution based on 486 not 'unstable' radio sources with z<1 and the third solution based on 582 not 'unstable' radio sources with z>1. Thus, the first solution includes more sources and observations. On the other hand, some of the sources are not astrometrically stable, therefore, the harmonic estimates for the first solution can be corrupted.

Table 1 presents estimates of the first degree spherical harmonics for three solutions. The estimates of the first degree electric spherical harmonics are stable with respect to change of the reference radio sources. The magnitudes of the differential secular aberration vector lie within theirs standard deviation errors. The first solution in Table 1 provides better formal statistics. However, the direction of this vector in the first solution is different from two other solutions, presumably, due to the effect of the radio sources structure instability.

| Solution | All radio sources | z<1 no 'unstable' | z>1 no 'unstable' |
|---|---|---|---|
| Secular aberration drift ($\mu$as/year) | 25.1 +/- 1.1 | 25.0 +/- 2.1 | 24.3 +/- 3.2 |
| Right Ascension | 263° +/- 3° | 269° +/- 5° | 278° +/- 6° |
| Declination | 20° +/- 3° | 45° +/- 7° | 40° +/- 8° |

Table 1. Estimates of the vector spherical harmonics l=1 for different sets of the reference radio sources

Table 2 shows the estimates of l=1 and l=2 spherical harmonics. The first solution was based on all radio sources observed in VLBI programs. The magnitude of the differential secular aberration vector reduces by 40 per cent if the second degree harmonics are added. However, this vector direction does not change significantly. The estimates of the second degree spherical harmonics are statistically significant. Some of them increase if the first degree spherical harmonics are not estimated. Uneven distribution of the reference radio sources results in high correlation among the l=1 and l=2 harmonics (up to 0.8). Until more radio sources in the southern hemisphere are observed, it is necessary to explore whether higher degree harmonics are significant or not.

| Parameter | (l,m) | l=1 only | l=1 and 2 | l=2 only |
|---|---|---|---|---|
| Secular aberration drift ($\mu$as/year) | | 25.1 +/- 1.1 | 15.6 +/- 2.1 | |
| Right ascension Declination | | 263° +/- 3° <br> 20° +/- 3° | 279° +/- 6° <br> 26° +/- 15° | |
| Electric harmonics ($\mu$as/year) | (2,0) <br> (2,1) <br> (2,-1) <br> (2,2) <br> (2,-2) | | 0.5 +/- 1.8 <br> -5.4 +/- 0.9 <br> -3.7 +/- 0.8 <br> 4.5 +/- 0.6 <br> 0.5 +/- 0.5 | 3.5 +/- 1.0 <br> -11.3 +/- 0.6 <br> -3.2 +/- 0.6 <br> 4.6 +/- 0.6 <br> 0.7 +/- 0.5 |
| Magnetic harmonics ($\mu$as/year) | (2,0) <br> (2,1) <br> (2,-1) <br> (2,2) <br> (2,-2) | | -5.9 +/- 0.8 <br> 3.2 +/- 0.7 <br> -9.7 +/- 0.7 <br> 2.4 +/- 0.5 <br> 1.4 +/- 0.5 | -5.6 +/- 0.8 <br> 2.7 +/- 0.6 <br> -13.1 +/- 0.6 <br> 2.3 +/- 0.6 <br> 1.4 +/- 0.5 |

Table 2. Estimates of the vector spherical harmonics for different combinations of parameters; l=1 only: l=1 and l=2; l=2 only. All radio sources were used as reference ones.

CONCLUSION

The systematic effects in the reference radio source proper motion have been indicated from global analysis of geodetic VLBI data. The first and second degree harmonic estimates were statistically significant. Disproportionate reference radio source distribution around the sky causes high correlation between these parameters and can result in a bias of the harmonic estimates. More astrometric VLBI observations in the southern hemisphere should be done in order to reduce the mutual correlation.

Acknowledgments:

I am thankful to S. Klioner, M. Eubanks, M. McClure, V. Vityazev, D. Jauncey, V. Zharov, D. MacMillan and N. Brown for fruitful discussion in the preparation of the manuscript.
The paper is published with the permission of the CEO, Geoscience Australia.